\documentclass[twocolumn,showpacs,amsmath,amssymb,prb]{revtex4}

\usepackage{graphicx}
\usepackage{bm}
\usepackage{verbatim}

\begin{document}

\title{Physics of Proximity Josephson Sensor}
\author{J. Voutilainen}
\author{M. A. Laakso}
\author{T. T. Heikkil\"a}
\affiliation{Low Temperature Laboratory, Aalto University School of Science and Technology, P.O. Box 15100, FIN-00076 AALTO, Finland}

\date{\today}

\begin{abstract}
We study the proximity Josephson sensor (PJS) in both bolometric and calorimetric operation and optimize it for different temperature ranges between 25~mK and a few Kelvin. We investigate how the radiation power is absorbed in the sensor and find that the irradiated system is typically in a weak nonequilibrium state. We show in detail how the proximity of the superconductors affects the device response: for example via changes in electron-phonon coupling and out-of-equilibrium noise. In addition, we estimate the applicability of graphene as the absorber material.
\end{abstract}


\maketitle

\section{Introduction}

As a class of high sensitivity infrared detectors, superconducting sensors operating at cryogenic temperatures are increasingly popular. For different design choices of the device, superconductivity guarantees greatly reduced heat leakage,\cite{KaraJAP, nature} drastic drop of resistance at the critical temperature\cite{Karasik} or negligible Johnson noise when operating in dissipationless regime.\cite{KID} The detector characterizing noise equivalent performance (NEP) values are thus expected to be as small as $10^{-19}$~W$/\sqrt{\rm Hz}$ at $T\approx 1$~K for the state-of-the-art detectors.\cite{Sergeev} Depending on whether the sensor operates in continuous or single-photon detection mode, it can be used in astrophysical space applications, security applications, spectroscopy of ultrafast quantum phenomena, optical communications, quantum cryptography or fast digital circuit testing.

There exist several measures to characterize the performance of a detector. Furthermore, their relative importance depends on the specific application in question. The requirements for a good detector include low noise, sufficient speed and good impedance matching between the receiver antenna and the detector element. Different detectors such as transition edge sensors and kinetic inductance detectors excel in different areas and operation regimes. When considering any one particular high-end application, the choice of the detector needs to be determined on a case-by-case basis, depending on the qualities and properties required.

Very recently, a model for a mesoscopic kinetic-inductance detector has been proposed.\cite{KID} This device is based on the coherent superconducting proximity effect, in contrast to some other superconductor--normal-metal--superconductor (SNS) detectors,\cite{SNSbolo1, SNSbolo2} where the superconductors are only used to block the energy outdiffusion from the radiation absorber and transport between the superconductors remains incoherent. The \emph{Proximity Josephson Sensor} (PJS) under consideration consists of a long SNS junction (length larger than the superconducting coherence length), where the temperature in the normal island changes under irradiation. This change in temperature causes an exponential change in the Josephson critical current between the two superconductors, which, in turn, greatly modifies the kinetic inductance of the sensor. This can be read with high accuracy using for example SQUID or resonant circuit based readout methods. As a result, performance values such as NEP $\sim 7\times 10^{-20}$~W/$\sqrt{\rm Hz}$ ($T=200$~mK) and signal-to-noise ratio $\sim 10^2$ ($f\sim$~THz) are expected, when the detector is operated in continuous and single-photon modes, respectively. In addition to favorable performance expectations, the advantages of this type of a detector include versatility, i.e., the fact that the device can be easily fabricated to optimize detection at the chosen operating temperature. Moreover, the response can be measured from a well-defined and spatially small region where temperature changes. The smallness of this region is relevant since the sensitivity depends on electron-phonon interaction, the strength of which is proportional to the volume. Finally, it is possible to get a good understanding of the physics of the device, as we show below.

\begin{figure}
\centering
\includegraphics[width=\columnwidth]{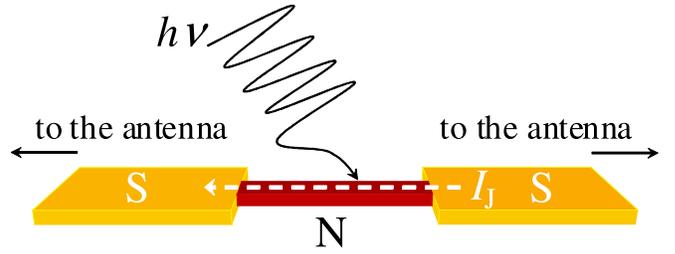}
\caption{(Color online) Illustration of the detector with normal (N) and superconducting (S) components. Incident radiation is coupled to the normal-conducting absorber via the antennas in contact to the superconductors.}\label{fig:dev}
\end{figure}

Here we detail the proposition made in Ref.~\onlinecite{KID}, formulate a model to explain the process of radiation coupling to the sensor and also take a closer look on how the proximity effect induced in the normal metal by the superconductors affects the performance of the detector. A number of distinguishable effects are due to the proximity induced minigap in the quasiparticle excitation spectrum of the normal island. Some examples of these are the reduced electron-phonon scattering and the reduced electron heat capacity. While considering the underlying physics of the detector, we also take on a practical approach to determine the required operating conditions more closely than has been done in Ref.~\onlinecite{KID}. The analysis contains some aspects general to all detectors with a similar hot-electron mechanism, especially when we study the coupling of the incident radiation to the absorber and the mechanism by which the radiation power is converted to a thermal signal. When the operating conditions are varied, different physical realizations in terms of detector materials may become favourable and therefore we also consider whether graphene could be used as the radiation absorber. The potential benefits in this case stem from the low volume and the subsequent decrease in intrinsic thermal noise of the two-dimensional sheet of carbon.

\section{Detector}
\subsubsection{Basic operating principle}
The detector consists of a diffusive metal wire (N) connected to two superconductors (S) through transparent contacts, as portrayed in Fig.~\ref{fig:dev}. Due to the proximity effect, superconducting correlations penetrate the normal wire and a steady supercurrent can be sustained between the superconducting terminals, in the absence of an external voltage. The maximum value of this supercurrent, $I_c$, is strongly dependent on the electron temperature $T_e$ in the wire in the temperature range $E_{\rm Th}<k_BT_e\ll\Delta$, where $\Delta$ is the energy gap in S and $E_{\rm Th}=\hbar D/l^2$ is the Thouless energy of the N wire with diffusion coefficient $D$. In short, incoming radiation increases electron temperature in the absorber, this increase in temperature makes the Josephson critical current $I_c(T_e)$ decrease, and the decrease in $I_c$ translates to a change in the kinetic inductance of the system, $L_k\sim 1/I_c$. The change in $L_k$ can then be measured by one of the standard techniques discussed below.

Since the detector responds to incident radiation with elevation of temperature, it is essential that the temperature range where the response is sufficiently large to be measured is within reasonable limits. In the diffusive wire, this range is determined by the energy minigap $E_g(\phi)$ induced in the wire due to the proximity effect. This minigap depends on the phase difference $\phi$ between the superconductors and for $\phi=0$ it obtains its maximum $E_g(0)=3.1E_{\rm Th}$.\cite{minigap} When the wire is long enough so that the overlap of S order parameters in the midpoint is small, $E_{\rm Th}\ll\Delta$ and the operating regime of the detector is determined to be well below the critical temperature of the superconductors. On the other hand, strong superconductors are needed to prevent outdiffusion of excited electrons into the leads before they relax. This outdiffusion grows larger with larger $E_{\rm Th}$ which is proportional to the diffusion constant $D$. For small $D$, the critical current becomes lower and consequently the operating temperature decreases. The exact regime depends on the measurement accuracy for the supercurrent and the desired frequency range of operation.

\subsubsection{Detailed discussion}\label{sec:deta}
The remainder of this section is devoted to a more detailed description of the operating procedure. Here and below, we strive for using simple analytical expressions for the essential physical properties of the detector. Inhomogeneous superconductivity in diffusive conductors can be described with the Usadel equation.\cite{usadel} To describe the proximity induced effects, we solve the Usadel equation numerically and calculate the response coefficients such as heat conductance between electrons and phonons from the solutions.\cite{weaklinkrev, usadelrev} This is required, for example, to determine the modified density of states for the absorber material. We then write down and use approximate expressions drawn from these exact numerical results. Secondly, we use a model where the effect of the electromagnetic radiation coupling to the absorber is assumed to show as an increase in the electron temperature of the absorber. The question of radiation coupling to superconducting elements is a long-standing problem which we unfortunately cannot answer in detail in the course of this discussion. Nevertheless, we expect that the result of this coupling can be described by a hot-electron model in our case, where the frequency of the incident photons $\nu\sim$~THz is much larger than the frequency corresponding to the edge of the energy minigap in the absorber $E_g/h\sim$~5~GHz.\cite{wavecoupl} We detail our reasoning in Sec.~\ref{sec:rad}.

\begin{figure}[htb]
\centering
\includegraphics[width=\columnwidth]{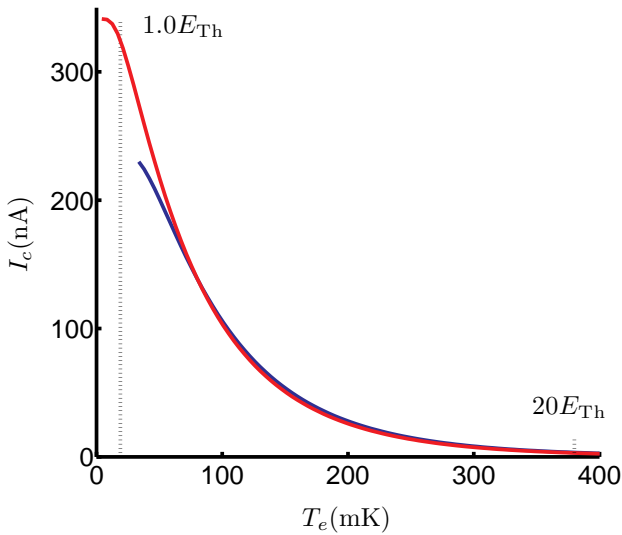}
\caption{(Color online) Critical Josephson current $I_c$ as a function of electron temperature $T_e$. We use ratio $\Delta/E_{\rm Th}=830$ for $E_{\rm Th}/k_B=19$~mK, corresponding to Nb energy gap $\Delta=16$~K$\times k_B$.}\label{fig:IcT}
\end{figure}

The critical Josephson current for a diffusive wire can be determined from the Usadel equation. For $E_{\rm Th}\ll k_BT_e\ll\Delta$, it is\cite{Ic}
\begin{equation}\label{eq:Ic}
\begin{split}
I_c&=\frac{64\pi k_BT_e}{(3+2\sqrt 2)eR_N}\sqrt{\frac{2\pi k_BT_e}{E_{\rm Th}}}\exp{\left(-\sqrt{\frac{2\pi k_BT_e}{E_{\rm Th}}}\right)}\\
&=\frac{64\pi k_BT_e}{(3+2\sqrt 2)eR_N}\frac{l}{\xi_N}\exp{\left(-\frac{l}{\xi_N}\right)}.
\end{split}
\end{equation}
Here, we have defined $\xi_N\equiv\sqrt{\hbar D/2\pi k_BT_e}$, and the normal state resistance of the wire, $R_N=\rho l/A=l/N_Fe^2DA$, is expressed in terms of the wire resistivity $\rho$, wire cross section $A$, and normal-state density of states $N_F$ at the Fermi level, in addition to the quantities already defined. The numerical solution for the critical current as a function of electron temperature together with the analytical approximation of Eq.~\eqref{eq:Ic} are shown in Fig.~\ref{fig:IcT} for some typical values of material parameters to be defined below.
The regime of strongest response can be seen to come about at temperatures of a few $E_{\rm Th}$ and while $I_c$ saturates at low temperatures, we see that values of $k_BT_e\gtrsim E_{\rm Th}$ are already sufficient for a sizable response $\partial I_c/\partial T_e$. This sets the lower limit for operation. The upper limit is set by the minimum measurable critical current, and here we assume this to be of the order of a few nanoamps corresponding to some tens of nH in the inductive readout.~\cite{Pertti}. In Fig.~\ref{fig:IcT}, critical current of 3~nA is obtained at temperatures of roughly 400~mK, corresponding to $20E_{\rm Th}$. It is notable that if the linearity of response is the preferred characteristic, the detector should be operated at a narrow temperature range of $\delta T_e\sim E_{\rm Th}$ within the chosen operating point.

Changes in $I_c$ are readily reflected in the kinetic inductance for a Josephson weak link, which is defined as $L_k=\hbar/2e\left(\partial I_S/\partial\phi\right)^{-1}\approx\hbar/2e({\rm cos}\phi)^{-1}$. The current-phase relation for the supercurrent is typically approximately sinusoidal, $I_S=I_c\sin(\phi)$. The most feasible value for $\phi$ to be used in operation of the sensor needs to be determined in the experiments and it depends on the measurement accuracy for the supercurrent. The readout can be performed by using a SQUID or probing the resonant frequency of an $LC$ resonator where the SNS junction provides the inductance. Specifically, in the previous case the magnetic flux noise can be as small as $10^{-7}\,\Phi_0/\sqrt{\rm Hz}$ ($\Phi_0$ is the magnetic flux quantum)\cite{squidsensis} so that the contribution from the readout can become 1-2 orders of magnitude smaller than the contribution from intrinsic sources and therefore negligible.\cite{KID} The sensitivity of the detector is then ultimately determined by the intrinsic sources, i.e., energy fluctuations in the absorber.

The response to the incoming radiation and the corresponding raise in temperature is mostly dependent on the volume $\Omega$ of the element absorbing the radiation energy and the degree of energy relaxation in the electron subsystem. We do not exclude the possibility of thermal transport into the leads but it is preferable to design the detector in a way to prevent the excess energy in the electron system escaping the island. Then, the electron-phonon coupling becomes the dominant mechanism for energy relaxation and this is in the normal state characterized by the electron-phonon coupling constant $\Sigma$. The size of the detector is essential in three ways. First, as shown below, the intrinsic noise of the sensor scales with respect to the volume of the radiation absorbing element and secondly, a small absorber responds more strongly to small incoming quanta of energy in terms of elevated temperature. Moreover, the normal metal bridge has to be short enough to be able to support measurable supercurrent. Thirdly, however, the detector cannot be too short or the outdiffusion rate of energy grows considerably. This means that typical scales lie near one micron for detector length and range from tens to hundreds of nanometers for cross-sectional dimensions. In addition to the absorber and the superconducting leads coupled to it, a separate antenna is required to couple the incoming radiation to the actual SNS sensor. The input impedance of the N wire can be varied to match the antenna impedance resulting in a good quantum efficiency. Most notably, a typical resistance of the N bridge is close to 50~$\Omega$ which usually gives the best matching.

\subsection{Coupling to radiation}\label{sec:rad}
The radiation detector studied in this work is based on a small electronic system absorbing the radiation. In such systems even a weak perturbation may give rise to a notable change in the state of the system. As shown below, in some cases this state cannot entirely be described via a simple increase of temperature. In the following, we aim to answer the three questions: (i) how is the energy distributed inside the absorber after the initial excitation, (ii) to which extent can this energy absorption be described by an increase in the electron temperature and (iii) how does the energy finally escape the system. A proper theory of the effects of radiation in a coherent SNS system for the frequencies of the order of $E_{\rm Th}$ is quite complicated\cite{wavecoupl}. However, for frequencies $h\nu>2\Delta$ we expect any coherent effects to be small.\cite{Pauli} Therefore, at frequencies close to terahertz it is justified to assume that coupling to the proximity device is similar to coupling to a normal metal.

To address the question of energy distribution and escape, we formulate a model based on the Boltzmann equation in a steady state,
\begin{equation}\label{eq:processes}
\begin{split}
0=\frac{\partial f(\epsilon)}{\partial t}=&I_{\rm rad}(\epsilon,\nu)+I_{\rm e-ph}(\epsilon)+I_{\rm e-e}(\epsilon)+\\
&\frac{8}{\tau_D}(f(\epsilon)-f_{\rm eq}(\epsilon,\;T_{\rm bath}))\Theta(|\epsilon|-\Delta),
\end{split}
\end{equation}
where the electron energy distribution function at energy $\epsilon$, $f(\epsilon)$, is modified from its equilibrium Fermi-Dirac form $f_N(\epsilon,\;T_{\rm bath})$ due to incident radiation of frequency $\nu$. The collision integrals $I_{\rm rad}(\epsilon,\nu),\;I_{\rm e-ph}(\epsilon),\;I_{\rm e-e}(\epsilon)$ describe the interaction between the electron having energy $\epsilon$ and the radiation field, absorber phonons and other electrons, respectively.\cite{RMP} The radiation collision integral is composed of all the processes corresponding to absorption and emission of a photon with frequency $\nu$,
\begin{equation*}
\begin{split}
I_{\rm rad}=&\frac{P_{\rm opt}}{(h\nu)^2}(f(\epsilon+h\nu)(1-f(\epsilon))+f(\epsilon-h\nu)(1-f(\epsilon))\\
&-f(\epsilon)(1-f(\epsilon+h\nu))-f(\epsilon)(1-f(\epsilon-h\nu)))\\
=&\frac{P_{\rm opt}}{(h\nu)^2}(f(\epsilon+h\nu)+f(\epsilon-h\nu)-2f(\epsilon)),
\end{split}
\end{equation*}
and the incident optical power is $P_{\rm opt}$. In the absence of coherent corrections the prefactor $P_{\rm opt}/(h\nu)^2$ is independent of energy. For $|\epsilon|>\Delta$, the possibility of the quasiparticle escaping to the superconducting leads is described by the diffusion term with the diffusion time $\tau_D=E_{\rm Th}/\hbar$\; and Heaviside step function $\Theta(|\epsilon|-\Delta)$. The latter accounts for the lack of heat diffusion into the superconductors for energies below $\Delta$. The origin of relaxation time for diffusion, $\tau_D/8$, is easily understood by noting that the quadratic length dependence of $\tau_D$ gives a factor 4 for diffusion length $L/2$ from the midpoint of the wire and since there are two possible directions for diffusion, the escape time is reduced from $\tau_D$ by the factor 8. In the superconductor at $|\epsilon|>\Delta$, the quasiparticles are assumed in equilibrium at the bath temperature with distribution $f_S(T_{\rm bath})$. The total power corresponding to each of these processes is obtained by multiplying by $\epsilon N_F$ and integrating over energy.

We solve Eq.~\eqref{eq:processes} to observe that the response of the system to incident radiation is almost thermal at typical materials parameters and a wide range of input powers $P_{\rm opt}$ and frequencies $\nu$. This means that the electron distribution function $f(\epsilon)$ of the irradiated system can, to a good accuracy, be approximated by an equilibrium Fermi function with a certain electron temperature $T_e$ determined by input power and some other parameters. Consequently, the supercurrent response is independent of the fine details of the nonequilibrium. However, $T_e$ cannot be obtained by a simple quasiequilibrium heat balance between the radiation power and the electron-phonon coupling. In most cases, this means that some fraction of the optical power flows out from the system in a way that it does not contribute to equilibrium heating of electrons and $T_e<T_e^0$, where $T_e^0$ is the resulting temperature in the case where an equilibrium system is heated with power $P_{\rm opt}$ and the energy flows fully into the phonons, as assumed for example in Ref.~\onlinecite{KID}. There are two distinct reasons for this loss of power. Below, we detail these processes.

\begin{figure}[htb]
\centering
\includegraphics[width=\columnwidth]{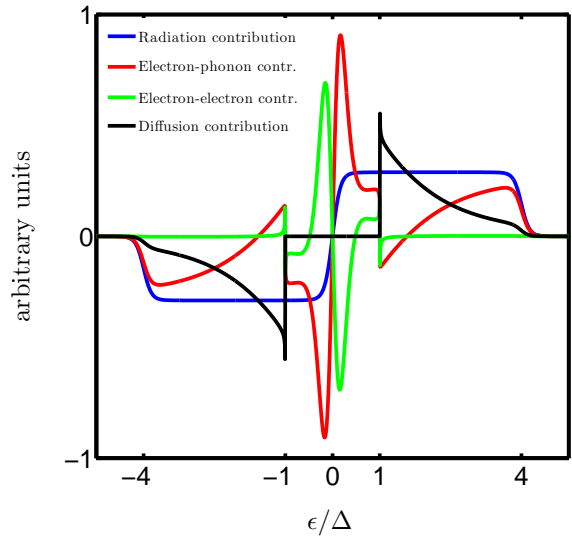}
\caption{(Color online) Relative contributions to the electron energy balance of the system defined in Eq.~\eqref{eq:processes}. At all energies $\epsilon$, the magnitude of the input (radiation contribution) must equal to the sum of the magnitudes of the other three processes which contribute to energy relaxation. The result is obtained for a typical set of operating and materials parameters and, most notably, $h\nu=4\Delta$ and $k_BT_e\approx 0.09\Delta$.}\label{fig:integrals}
\end{figure}
The different contributions to electron energy balance defined in Eq.~\eqref{eq:processes} are shown for our example system in Fig.~\ref{fig:integrals}. Since electron-electron interaction only mixes electrons with different energies and does not remove energy from the system, the corresponding power is zero. Due to the conservation of energy, the power corresponding to the radiation contribution then has to equal the integral over the sum of the electron-phonon integral and the diffusion part weighed by $\epsilon N_F$. The relative magnitude of these two parts gives the fraction of power which either escapes to the leads or to the phonons via the electron system in the absorber while only the latter results in a detector response by the means of an increased temperature. This is first of the two causes for reduced thermal power absorption and it is the most prominent for frequencies $h\nu\gtrsim\Delta$. When $h\nu\gg\Delta$, most of the radiation power is distributed at high energies where strong electron-phonon relaxation dominates outdiffusion from the island and the relative fraction of escaping power becomes small. This may be quantified within the the hot-electron assumption and relaxation time approximation for electron-phonon interaction, when $I_{\rm e-ph}=(f(\epsilon,\;T_e)-f_N(\epsilon,\;T_{\rm bath}))/\tau_{\rm e-ph}(\epsilon)$. As determined numerically, electron-phonon interaction is the dominant factor for inelastic relaxation and we may simplify Eq.~\eqref{eq:processes} considerably. Since $h\nu\gg k_BT_e$, $I_{\rm rad}$ is almost constant, $I_{\rm rad}={\rm sign}(\epsilon)P_{\rm opt}/(h\nu)^2\Theta(h\nu-|\epsilon|)$, for energies below the radiation frequency. The corresponding power is divided between outdiffusion and inelastic (electron-phonon) relaxation with ratio
\[
r_1\equiv\frac{P_{\rm e-ph}}{P_{\rm tot}}=\frac{2\int_0^{h\nu}\tau_{\rm e-ph}(\epsilon)^{-1}\epsilon\,d\epsilon}
{2\int_0^{h\nu}(8\tau_D^{-1}\Theta(|\epsilon|-\Delta)+\tau_{\rm e-ph}(\epsilon)^{-1})\epsilon\,d\epsilon}
\]
for the power absorbed in the detector. The energy-dependent electron-phonon time $\tau_{\rm e-ph}(\epsilon)=72\zeta(5)k_B^5N_F/(\Sigma\epsilon^3)$\cite{RMP, Rammer}, where $\zeta(x)$ is the Riemann zeta function, $\zeta(5)\approx 1.0369$, and thus
\begin{equation}\label{eq:r1}
r_1=\left(1+\frac{1440\zeta(5)k_B^5N_F}{\tau_D\Sigma}\frac{(h\nu)^2-\Delta^2}{(h\nu)^5}\right)^{-1}.
\end{equation}
The ratio approaches unity for $h\nu\gg\Delta$ and has its minimum at $h\nu\approx 1.3\Delta$, independent of other parameters.

The second cause of reduced thermal absorption is due to the relative weakness of electron-electron interaction. It turns out that for a typical small diffusive wire the electron-electron interaction becomes negligible compared to the electron-phonon interaction at energies that are generally smaller than the operating scales we are interested in, namely $\Delta$. This is in contrast to some earlier assumptions about the source of relaxation in similar hot-electron bolometers.\cite{KaraJAP} The absence of electron-electron collisions changes notably the relaxation and distribution of the incident radiation power as electron-phonon collisions are required to balance the radiation power especially at high energies. In Fig.~\ref{fig:integrals} this is seen as a deviation of the electron-phonon contribution from its quasiequilibrium form where a sharp peak at low energies decays with width determined by the electron temperature. Instead, in the radiation-induced nonequilibrium state electron-phonon interaction is strong also at energies $\epsilon\gg T_{\rm eff}$. The resulting increase in the electron temperature is smaller than expected from simple heat balance. We attribute this to the fact that electron-phonon relaxation at high energies is much faster than at low energies, resulting in a large heat conductance for the nonequilibrium component excited by the radiation power. Numerically, we observe that the fraction of incident power converted to thermal response (in the absence of diffusion) $r_2\sim 1/(h\nu)$ at high frequencies. Compared to the frequency dependence, the depence of $r_2$ on the radiation power is relatively weak.

\begin{figure}[htb]
\centering
\includegraphics[width=\columnwidth]{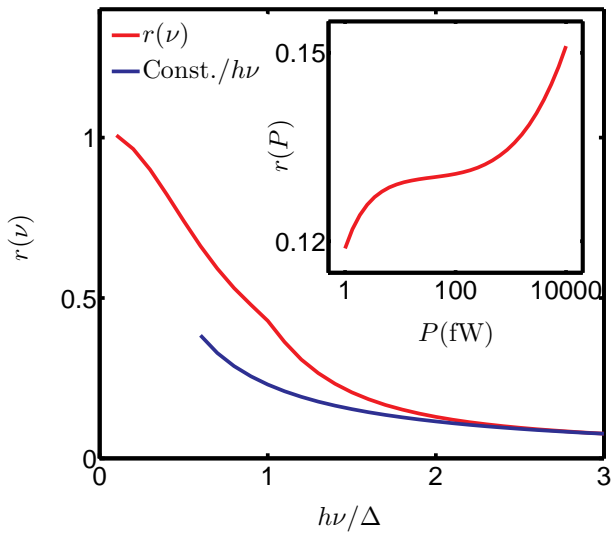}
\caption{(Color online) Fraction $r(\nu,P)$ of incident radiation power converted into thermal response in a PJS detector. The results are given as a function of $\nu$ for $P=10$~fW and as a function of $P$ for $h\nu=2\Delta$ (inset). The curves are obtained for a typical set of materials parameters defined in Sec.~\ref{sec:par}.}\label{fig:r}
\end{figure}
The effects described by $r_1$ and $r_2$ are not completely independent. The results for the case where both the outdiffusion ($r_1$) and the distribution of energy through inelastic relaxation $r_2$ are taken into account are shown in Fig.~\ref{fig:r} for our example system. For small $h\nu$, the frequency dependence is complicated due to the competition between the two effects but when $h\nu\gg\Delta$, the loss of power follows the $1/\nu$-law.

\begin{figure}[htb]
\centering
\includegraphics[width=\columnwidth]{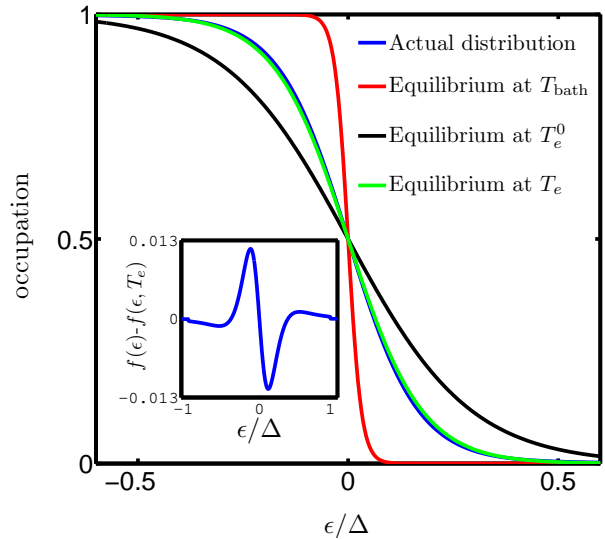}
\caption{(Color online) Nonequilibrium distribution resulting from irradiation and a set of relevant equilibrium distributions. Notice that the curves for the actual nonequilibrium distribution $f(\epsilon)$ and the equilibrium distribution $f(\epsilon,T_{\rm eff})$ used to approximate this are nearly on top of each other. The difference between these two is magnified in the inset. The figure is obtained for the same set of parameters used in Fig.~\ref{fig:integrals}. In this case the input power $P_{\rm opt}$ is large so that the bath temperature $T_{\rm bath}$ is small compared to the final $T_e$. Here $r\approx 0.08$, so $T_e$ is also clearly smaller than the temperature of an equivalent equilibrium system heated with the same power, $T_e^0$.}\label{fig:fstudy}
\end{figure}
In Fig.~\ref{fig:fstudy}, the nonequilibrium electron distribution is compared to related equilibrium distributions. The essential difference between the nonequilibrium distribution and its quasiequilibrium approximation is the slight increase in the fraction of electrons at high energies. While this changes essentially the electron-phonon collision integral in Eq.~\eqref{eq:processes} to balance the high-energy radiation excitations even for $h\nu\gg\Delta$, it has no relevant effect on the supercurrent response of the system. We therefore use the quasiequilibrium assumption with electron temperature $T_e$ below as a good approximation.

\subsection{Parameters and conditions}\label{sec:par}
To quantify the discussion, we assume leads of superconducting Nb and a wire of normal-conducting Ag (also Cu is a viable alternative). As discussed in the previous section, a large superconducting energy gap $\Delta$ is needed to prevent energy outdiffusion from the absorber and to guarantee good quantum efficiency. The choice of superconductor is made solely with this in mind and it does not affect the performance in any other way since it is the minigap $E_g$ instead of $\Delta$ which determines the properties of the detector, as long as the condition $\Delta\gg k_BT_e,\,E_{\rm Th}$ is satisfied. The lead parameters are then $T_c\approx 9$~K, $\Delta=1.4$~meV, and for the wire, we assume length $l=1\;\mu$m, width $w=100$~nm, thickness $t=30$~nm, $N_F=10^{47}$~J$^{-1}$~m$^{-3}$, $D=25$~cm$^2$s$^{-1}$. This gives $\Omega=10^{-21}$~m$^3$, $R_N\approx 52\,\Omega$, $E_{\rm Th}\approx 1.6\,\mu$eV$=19$~mK$\times k_B$ and the proximity minigap in the wire at $\phi=0$ is $E_g\approx 60$~mK$=1.2$~GHz$\times h$. The limit $\Delta\gg E_{\rm Th}$ is well satisfied with $\Delta/E_{\rm Th}=830$. Furthermore, we use electron-phonon coupling coefficient $\Sigma=5\times 10^8$~Wm$^{-3}$~K$^{-5}$ characteristic for Ag.\cite{RMP} We use these values in the examples throughout the paper until Sec.~\ref{sec:opt}, where we study different detector setups.

While the acknowledged advantage in hot electron microbolometers is the small size of the absorbing element, resulting in increased responsivity, the important parameter in this particular design is the Thouless energy, $E_{\rm Th}=\hbar D/l^2$, determined by the diffusion constant $D$ and the length $l$ of the absorber. We note that it is precisely the large value of $E_{\rm Th}$ which differentiates PJS from the previous SNS-type detectors. For a detector of 1~$\mu$m in length, the coherent transport between superconductors persist up to temperatures comparable to the minigap $E_g\sim 3.1E_{\rm Th}\sim 60$~mK and as discussed above, exploiting the decay of this coherence is the integral part of the detector operation. Note that the operating range for the detector extends well beyond $E_g$ and can be determined from Fig.~\ref{fig:IcT}. In the SNS detectors of Refs.~\onlinecite{SNSbolo1} and \onlinecite{SNSbolo2}, the absorbers are roughly 5 times longer and, since $E_g\sim E_{\rm Th}\sim l^{-2}$, they reside in the incoherent regime for temperatures above 10~mK. Consequently, the sole purpose of the superconductors there is to prevent energy outdiffusion from the absorber. On the other hand, to observe strong proximity effect the absorber film needs to be sufficiently clean as well. For ultrathin silver samples with thickness of 50~nm, diffusivities of 200~cm$^2$/s have been measured,\cite{refD} but to minimize the noise of the detector, it is preferable also to minimize the thickness, even in the expense of diffusivity increased by the surface scattering. Our objective to strengthen the superconducting correlations by requiring clean samples is in contrast to the conventional approach in hot-electron detectors.\cite{KaraJAP} However, in contrast to Ref.~\onlinecite{KID} where $D=100$~cm$^2$ is used, we need to compromise in choosing weaker diffusivity to prevent absorbed energy from escaping to the superconductors. This corresponds to reduction of $E_{\rm Th}$ by a factor of 4 which consequently dulls the proximity effect, decreaces the critical current and shifts the operating regime of the detector to lower temperatures.

\section{Operation}\label{sec:oper}
The device can be designed to operate in two modes depending on the rate of the incident photons: continuous and single-photon detection. These are referred to as the bolometric and calorimetric modes, respectively.  Below, we study the system with particular focus on the effects that stem from the superconducting correlations induced in the absorber by the proximity effect. For comparison, we refer to the case where these correlations are taken into account as the \emph{coherent} case and the one, where the correlations are neglected, as the \emph{incoherent} case.

\subsection{Bolometer}\label{sec:bolo}
When the detector is subject to a continuous radiation with power $P_{\rm opt}$, its temperature is determined by the balance equation for incoming and outgoing powers,
\[
r(\nu)P_{\rm opt}\equiv P_T=\dot Q_{\rm e-ph}.
\]
Factor $r(\nu)$ accounts for the effects related to the coupling of radiation, discussed in Sec.~\ref{sec:par}, and for frequencies $h\nu>k_BT_e$ it is almost independent of temperature. The factor $r(\nu)$ may also contain effects such as possible impedance mismatch in calibration of the antenna circuit which determines the quantum efficiency with which the dissipation of the electromagnetic energy can be focused on the absorbing element N. After the incoming radiation energy is first absorbed in the electron system, the superconducting gap in the leads prevents the heat from escaping the island. The system is consequently relaxed by the means of energy flow from the electrons to the phonons acting as a heat bath. In the normal state, the heat flux is given by\cite{PisTto5}
\begin{equation}\label{eq:Q3D}
\dot Q_{\rm e-ph}^N=\Sigma\Omega(T^5_e-T_{\rm bath}^5),
\end{equation}
and the proximity effect can be described by using an approximate expression\cite{eph}
\begin{equation}\label{eq:Qappr}
\dot Q_{\rm e-ph}=\dot Q_{\rm e-ph}^Ne^{-T^*/T_e}.
\end{equation}
The exponential decay of the interaction is characterized by a constant, $T^*$, and it is due to the presence of the minigap $E_g(\phi)$ in the wire. In Ref.~\onlinecite{eph}, a good fit for $\dot Q_{\rm e-ph}$ is obtained in the case $\phi=0$ by using $T^*=3.7E_{\rm Th}/k_B$. The elevated temperature in the wire can then be determined by solving
\begin{equation}\label{eq:boloTe}
T_e^5=T_{\rm bath}^5+e^{T^*/T_e}\frac{P_T}{\Sigma\Omega}
\end{equation}
numerically.

\begin{figure}[htb]
\centering
\includegraphics[width=\columnwidth]{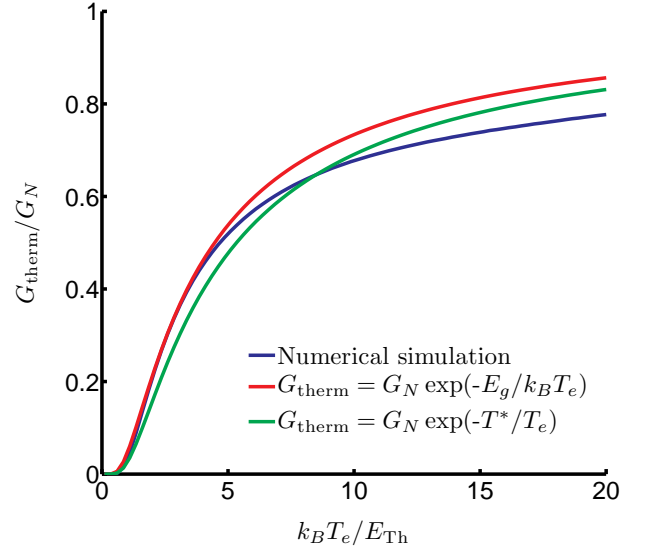}
\caption{(Color online) Comparison between numerical result for thermal conductance $G_{\rm therm}$ of the electron-phonon heat link and two exponential fitting curves: one valid to good degree of accuracy for $\dot Q$ (green) and one, where the suppression of $G_{\rm therm}$ is determined by the energy minigap $E_g$ (red). $G_{\rm therm}$ is given in proportion to the thermal conductance $G_N$ in the incoherent case.}\label{fig:G}
\end{figure}

The dominant contribution to the detector noise comes from the thermal fluctuations between the electron and phonon systems in the N bridge and we refer to this as the thermal fluctuation noise (TFN). Due to the superconducting gap, energy exchange between the N bridge and the leads becomes significant only at temperatures $k_BT_e\sim \Delta$. We first give a simple way to calculate the NEP of the bolometer in linear response, i.e., with respect to small deviations between $T_e$ and $T_{\rm bath}$. This is the usual procedure in determining the noise, valid when the input power is small and the system is close to equilibrium so that $T_e\approx T_{\rm bath}$. In equilibrium, the fluctuation-dissipation theorem states that NEP$_{\rm TFN}$ (directly related to the thermal noise correlator $\dot S_Q$) is given by
\begin{equation}\label{eq:NEP}
\begin{split}
{\rm NEP}_{\rm TFN}\equiv\sqrt{S_{\dot Q}}&=\sqrt{4k_BT_e^2G_{\rm therm}}\\
&\approx\sqrt{20\Sigma\Omega k_BT_e^6e^{-T^*/T_e}}\\
&={\rm NEP}_{\rm TFN}^{\rm N}e^{-T^*/2T_e}.
\end{split}
\end{equation}
Here the thermal conductance of the heat link is
\[
G_{\rm therm}=\frac{\partial\dot Q_{\rm e-ph}}{\partial T_e}=5\Sigma\Omega T_e^4e^{-T^*/T_e},
\]
and the effect of superconducting correlations is seen as reduction of noise from its normal state value ${\rm NEP}_{\rm TFN}^{\rm N}$ by an exponential factor.
We have also performed a numerical simulation to determine $G_{\rm therm}$ exactly and in Fig.~\ref{fig:G} we compare this to the two possible exponential formulas that can a priori be anticipated to determine the suppression of thermal conductance. We see that while the value $T^*=3.7E_{\rm Th}/k_B$ used to fit $\dot Q_{\rm e-ph}$ in Ref.~\onlinecite{eph} is adequate for our purposes in the range $E_{\rm Th}<k_BT_e<20E_{\rm Th}$, the fit is not optimal. Therefore we have also tried making the well-founded assumption that the suppression of $G_{\rm therm}$ is determined by the minigap, so that $G_{\rm therm}\sim\exp{\left(-E_g/k_BT_e\right)}=\exp{\left(-3.1E_{\rm Th}/k_BT_e\right)}$ (at $\phi=0$). In this case, the fit is improved at temperatures $T_e\lesssim 6E_{\rm Th}/k_B$.

Depending on the input optical power, the PJS can have $T_e$ markedly elevated from $T_{\rm bath}$ and the linear-response result given above is no more sufficient. In this case the TFN amplitude has to be calculated from the full analysis of the out-of-equilibrium response. We have  performed this calculation for the noise, yielding
\begin{equation}\label{eq:SQ}
\begin{split}
S_{\dot Q}=\int\,d\epsilon\,d\epsilon'\frac{\Sigma\Omega}{24\zeta(5)}(\epsilon-\epsilon')^4\,
{\rm sgn}\,(\epsilon-\epsilon')\\ 
\times K(\epsilon,\epsilon')\left\{n(\epsilon-\epsilon')f(\epsilon')(1-f(\epsilon))\right.\\
\left.+(1+n(\epsilon-\epsilon'))f(\epsilon)(1-f(\epsilon'))\right\},
\end{split}
\end{equation}
where the Riemann zeta function $\zeta(5)\approx 1.0369$ and $K(\epsilon,\epsilon')$ describes the effect of superconductivity, see~Ref.~\onlinecite{eph}. Equation~\eqref{eq:SQ} is valid in full nonequilibrium where the electron and phonon distribution functions $f(\epsilon)$ and $n(\epsilon)$ can be arbitrary, as long as the electron-hole balance is satisfied, i.e., $f(\epsilon)=1-f(-\epsilon)$. We will study the nonequilibrium corrections to these results in the future whereas here we use the hot-electron model where the electron and phonon systems can each be described by a thermal distribution with temperatures $T_e$ and $T_{\rm bath}$. The distribution functions are given by the Fermi function $f(\epsilon)=({\rm exp}(\epsilon/k_BT_e)+1)^{-1}$ and the Bose function $n(\epsilon)=({\rm exp}(\epsilon/k_BT_{\rm bath})-1)^{-1}$ for electrons and phonons, respectively. The structure of Eq.~\eqref{eq:SQ} can be understood by looking at the processes visualized on the second and third lines: they correspond to absorption and emission of a phonon, respectively. For noise, these processes, where heat flows either into or out of the N island, are additive and consequently show up as a simple sum. 

We first estimate analytically the high temperature behaviour of $S_{\dot Q}$ from Eq.~\eqref{eq:SQ}. This is done with the absorber in the incoherent limit, when the superconducting corrections to the absorber are neglected and $K(\epsilon,\epsilon')=1$. Equation~\eqref{eq:SQ} simplifies to
\begin{equation}\label{eq:bololimit}
\begin{split}
S_{\dot Q}=\frac{4\Sigma\Omega}{3\zeta(5)}k_BT_e^6 &\int dx \,x^5{\rm sgn}\,x\\
&\times\left[\coth{x}\coth{\left(\frac{T_e}{T_{\rm bath}}x\right)}-1\right],
\end{split}
\end{equation}
and, at the two opposite limits of low and high electron temperature, this yields:
\begin{subequations}\label{eq:Sab}
\begin{align}
S_{\dot Q}= &\;20\Sigma\Omega k_BT_e^6, \quad &T_e\approx T_{\rm bath}\label{eq:Sa}\\
S_{\dot Q}= &\;\frac{\zeta(6)}{\zeta(5)}10\Sigma\Omega k_BT_e^6\approx 9.8\Sigma\Omega k_BT_e^6, \quad &T_e\gg T_{\rm bath},\label{eq:Sb}
\end{align}
\end{subequations}
where the first line is in accordance with the linear-response result Eq.~\eqref{eq:NEP}. Moreover, both of these limiting values are in accordance with the results of Ref.~\onlinecite{Golwala} for TFN in a nonequilibrium setting with normal conductor.

\begin{figure}[thb]
\centering
\includegraphics[width=\columnwidth]{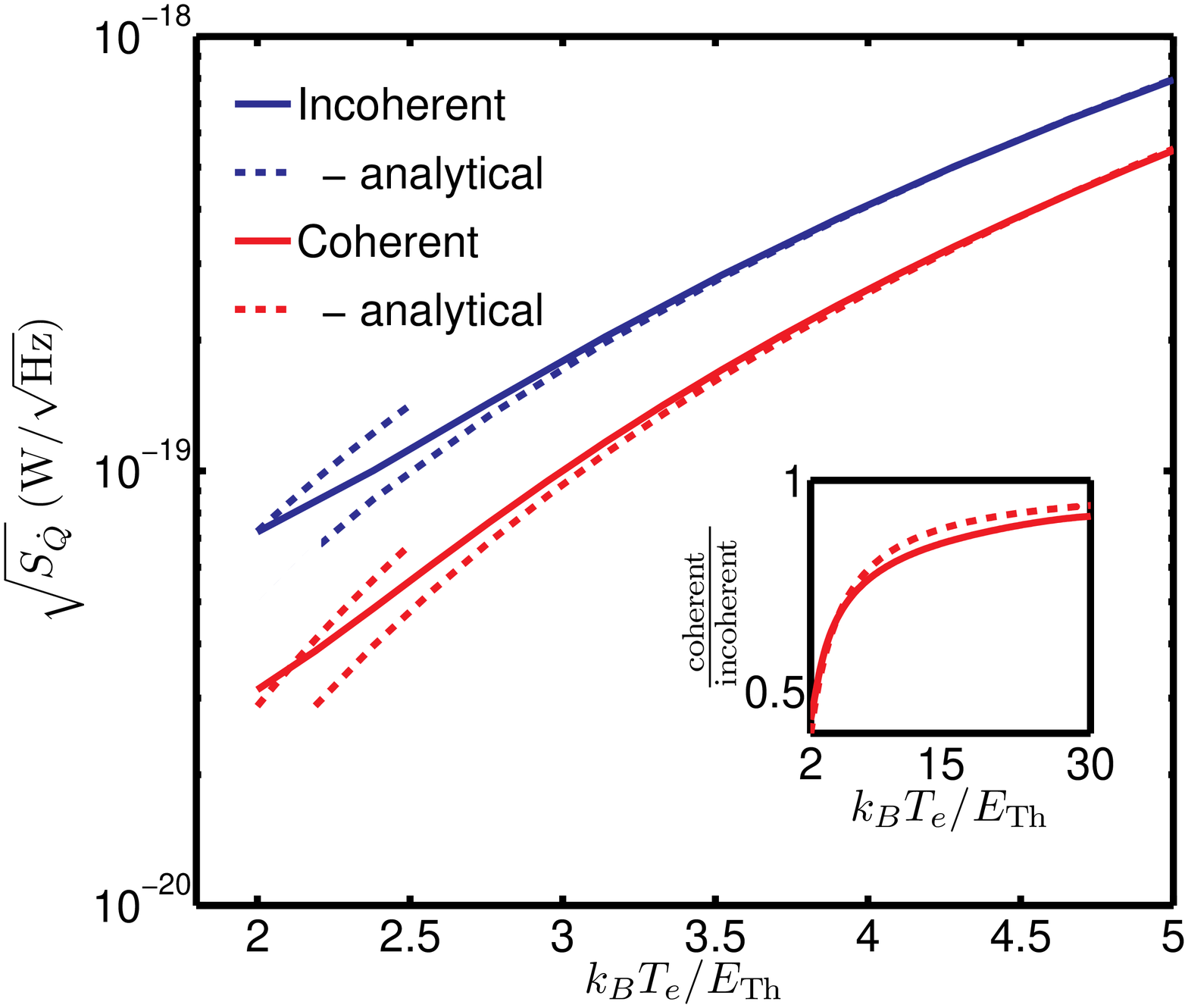}
\caption{(Color online) Nonequilibrium noise of the proximity Josephson sensor under heating ($k_BT_{\rm bath}=2E_{\rm Th}$), with (coherent - red) and without (incoherent - blue) the corrections from the proximity effect. We compare the exact results to the analytical approximations of Eqs.~\eqref{eq:Sab} (incoherent) and Eq.~\eqref{eq:NEP} (coherent) valid at the limits $T_e=T_{\rm bath}$ and $T_e\gg T_{\rm bath}$ (dashed lines). In the inset, we show how the noise of the coherent system can be approximated by the exponential law of Eq.~\eqref{eq:NEP} (dashed line) for a large temperature range.}\label{fig:SQ}
\end{figure}
We also solve Eq.~\eqref{eq:SQ} numerically to analyze the detector performance in the coherent limit, where the superconducting corrections are taken into account. The nonlinear noise described by Eq.~\eqref{eq:SQ} is shown in Fig.~\ref{fig:SQ} for our proximity SNS detector compared to the result in the absence of the superconducting proximity effect. Note that, due to the strong temperature dependence of $S_{\dot Q}$, the limit of high temperatures in Eq.~\eqref{eq:Sb} is achieved easily. Already at $T_e=2T_{\rm bath}$, the difference between the estimate of Eq.~\eqref{eq:Sb} and the exact value from Eq.~\eqref{eq:bololimit} is less than 2 percent. This means that we can safely use Eq.~\eqref{eq:Sb} as the upper limit for noise whenever the absorber is heated. In addition, as expected from Eq.~\eqref{eq:NEP}, we observe reduction in the noise $\sqrt{S_{\dot Q}}$ due to superconducting correlations. Nevertheless, in the operating range of the detector $k_BT_e > E_{\rm Th}$ this reduction is relatively small, namely a little more than a factor of 2 for our given values $k_BT_e=k_BT_{\rm bath}=2E_{\rm Th}$ as seen in the inset of Fig.~\ref{fig:SQ}. From the inset, we also see that the numerically calculated noise follows the exponential law of Eq.~\eqref{eq:NEP} with satisfactory accuracy. This, in turn, justifies the use of approximate expression introduced in Eq.~\eqref{eq:NEP} in our analysis. Below, we use a combination of Eqs.~\eqref{eq:NEP} and \eqref{eq:SQ} to give analytical estimates of the detector performance.

\begin{figure}[thb]
\centering
\includegraphics[width=\columnwidth]{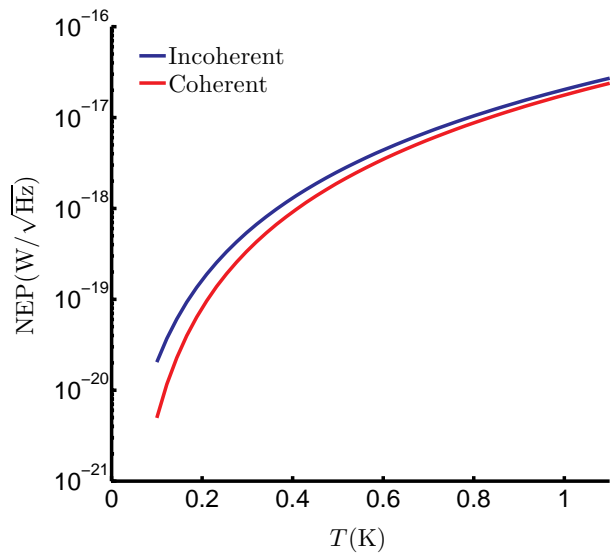}
\caption{(Color online) Base NEP (with negligible power load) for proximity and normal conductor as function of $T\equiv T_e=T_{\rm bath}$. The proximity effect decreases NEP by the factor $\sqrt{e^{3.7E_{\rm Th}/k_BT}}\approx 4$ in our example at $T=100$~mK.}\label{fig:NEP}
\end{figure}
Based on the limits of Eqs.~\eqref{eq:Sab}, we may also estimate the behavior of the detector at linear response, i.e., for small input power, and the corresponding NEP value for different $T_{\rm bath}$ is shown in Fig.~\ref{fig:NEP}. This is the base NEP. As seen from Eqs.~\eqref{eq:boloTe} and \eqref{eq:NEP}, the heating of the absorber due to the incoming optical power starts to make a comparable contribution to the noise when $P_T\sim\Sigma\Omega T_{\rm bath}^5e^{T^*/T_{\rm bath}}\approx 10^{-18}\,{\rm W}\ldots\, 10^{-12}\,{\rm W}$, in the temperature range 0.1~K~$\ldots$~1~K. Beyond this level of incident power, the noise becomes independent of $T_{\rm bath}$ and
\begin{equation}\label{eq:powerNEP}
\textrm{NEP}\sim\sqrt{10k_B}\sqrt[10]{\frac{P_T^6}{\Sigma\Omega}}\approx 2\times 10^{-10}\times P_T^{3/5},
\end{equation}
in dimensions of W$/\sqrt{\rm Hz}$. For an input power 5~pW, typical to balloon and ground-based astronomical telescopes, this translates to NEP $\sim 3\times 10^{-17}$~W/$\sqrt{\rm Hz}$.\cite{kuzmin} We stress that for this type hot-electron bolometer, the noise level is almost independent of any other parameters than the input power. A factor of 1000 change in the volume of the detector changes the noise only by a factor of 2. Only when the input power is below $\sim\Sigma\Omega T_{\rm bath}^5$, which corresponds to the intrinsic thermal fluctuations of the island, can the overall noise be reduced by decreasing the size of the absorber. When $P_T$ is larger so that dominant contribution to noise results from the radiation-induced heating, the noise can actually be reduced, albeit only slightly, by \emph{increasing} the size of the island.

From the above considerations, we determine the dynamic range of the detector, i.e., the range of input powers that can be detected with the sensor. The lower limit is given by NEP and the upper limit, proportional to NEP, results from the large increase in temperature $T_e$ when the sensor is subject to high level of incident radiation. The results of increasing $T_e$ are two-fold: firstly, the critical current is decreased below the level that can be detected by the readout circuit and secondly, the response $\partial I_c/\partial T_e$ becomes too small. The change in temperature is estimated by taking incident power equal to the NEP given in Eq.~\eqref{eq:NEP} multiplied by a factor $\alpha$, and using the result of Eq.~\eqref{eq:bololimit} at large $T_e$ so that
\begin{equation*}
\begin{split}
T_e&\approx\sqrt[5]{T_{\rm bath}^5+\frac{\alpha{\rm NEP(T_e=T_{\rm bath})}\sqrt{\delta f}}{\Sigma\Omega}}\\
&=T_{\rm bath}\sqrt[5]{1+\alpha\sqrt{\frac{20k_B\delta f}{\Sigma\Omega T_{\rm bath}^4}}},
\end{split}
\end{equation*}
where $\delta f$ is the frequency band used in the measurements. Again, using the values given in Sec.~\ref{sec:par} and assuming $\delta f=50$~Hz, $T_B=1$~K, $\alpha=10^6$, we obtain $T_e\approx 2.5T_{\rm bath}$, which can realistically be measured. If we reduce temperature to $T_{\rm bath}=0.1$~K, $T_e\approx 6.3T_{\rm bath}$ and for this, one needs to operate close to $T_{\rm bath}=E_{\rm Th}$ so that $T_e$ does not become too large. As the dynamic range is considered, it remains important to decide whether we wish to obtain linear output from the detector, so that only a restricted temperature range is available due to the exponential depence of $I_c$ at high temperatures (as in Fig.~\ref{fig:IcT}). The dynamic range is therefore dependent on how the detector is optimized but typical values of $\alpha=10^6$ are still well within reach.

\subsection{Calorimeter}
\begin{figure}[htb]
\centering
\includegraphics[width=\columnwidth]{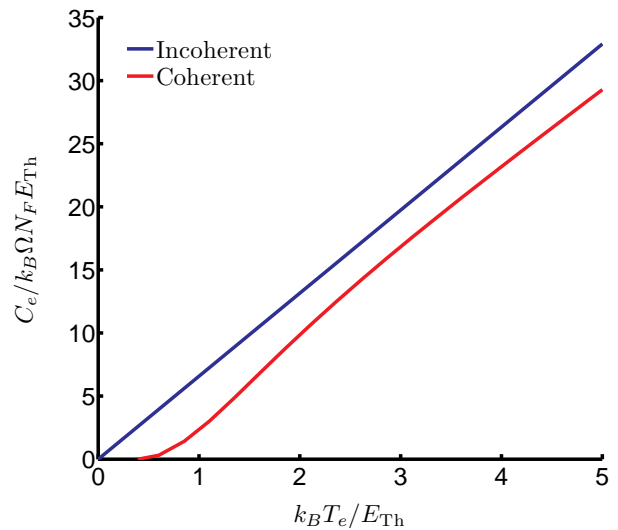}
\caption{(Color online) Comparison between the numerical result for the electronic heat capacity $C_e$ of the coherent (proximity) and the incoherent case.}\label{fig:C}
\end{figure}

In calorimetric mode, the rate of incident photons is much smaller than the rate of recovery for the detector. Consequently, photons are absorbed one at a time and after each absorption event, $T_e$ rises rapidly within the diffusion time $\tau_D=D/l^2$, and then decays back to the equilibrium value $T_{\rm bath}$ within time $\tau$, determined by the relaxation processes. We assume that the rise time of the pulse is very short, $\tau_D\ll\tau$, compared to the thermal time constant of the detector, which reads
\begin{equation}\label{eq:tau}
 \tau=\frac{C_e}{G_{\rm therm}}.
\end{equation}
The heat capacity of the electron system is given by
\begin{equation}\label{eq:C}
 C_e(T_e)=T_e\frac{\partial \mathcal{S}(T_e)}{\partial T_e},
\end{equation}
where the entropy\cite{C, S}
\begin{equation}\label{eq:S}
\begin{split}
 \mathcal{S}=&-2k_B\Omega\int d\epsilon\, N(\epsilon)\\
&\times\left\{f(\epsilon){\rm ln}\left[f(\epsilon)\right]+\left[1-f(\epsilon)\right]{\rm ln}\left[1-f(\epsilon)\right]\right\}.,
\end{split}
\end{equation}
and $f(\epsilon)=\left[1+\exp{\left(\epsilon/k_BT_e\right)}\right]^{-1}$. Without proximity effect, the density of states is independent of energy, $N(\epsilon)=N_F$, and $C_e^N=2\pi^2N_F\Omega k_B^2T_e/3$. The proximity effect modifies the density of states by opening the minigap $E_g\sim E_{\rm Th}\ll\Delta$.\cite{diffuDOS} As a result, the heat capacitance gets modified, but only sligthly, as shown in Fig.~\ref{fig:C}. The growth in $C_e$ is exponential when $k_BT_e<E_{\rm Th}$, after which $S$ saturates and $C_e$ becomes linear, close to the normal state value.

\begin{figure}[htb]
\centering
\includegraphics[width=\columnwidth]{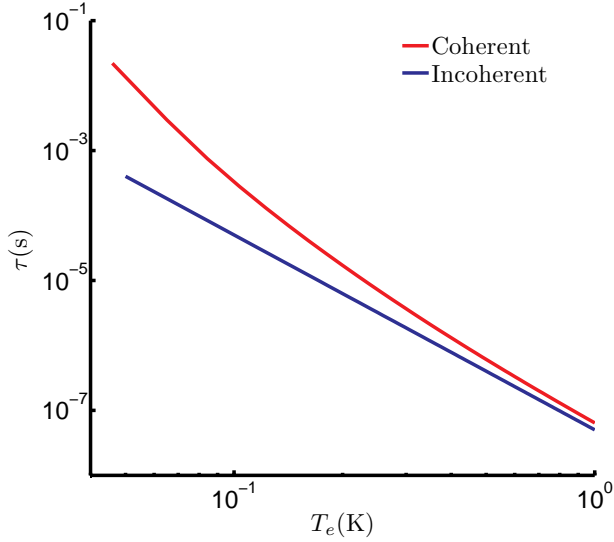}
\caption{(Color online) Time constant of the proximity Josephson sensor as a function of electron temperature on a log-log scale. The coherent corrections from the proximity effect (red) are compared to the incoherent result $\tau\sim T_e^{-3}$ (blue).}\label{fig:tau}
\end{figure}

Now that we have solved for the heat capacity $C$ and thermal conductance $G_{\rm therm}$ numerically, the time constant $\tau$ may be obtained from them using Eq.~\eqref{eq:tau}. Since the heat capacity as seen in Fig.~\ref{fig:C} is approximately linear, we may conclude that the temperature dependence of $\tau$ follows the law $\tau\sim e^{T^*/T}/T^3$. The exact results from numerical simulations are shown in Fig.~\ref{fig:tau} and from there, we see that $\tau_{\rm e-ph}\sim 10^{-3}\ldots 10^{-7}$~s in the range $T_e=0.1\ldots 1.0$~K.

\begin{figure}[htb]
\centering
\includegraphics[width=\columnwidth]{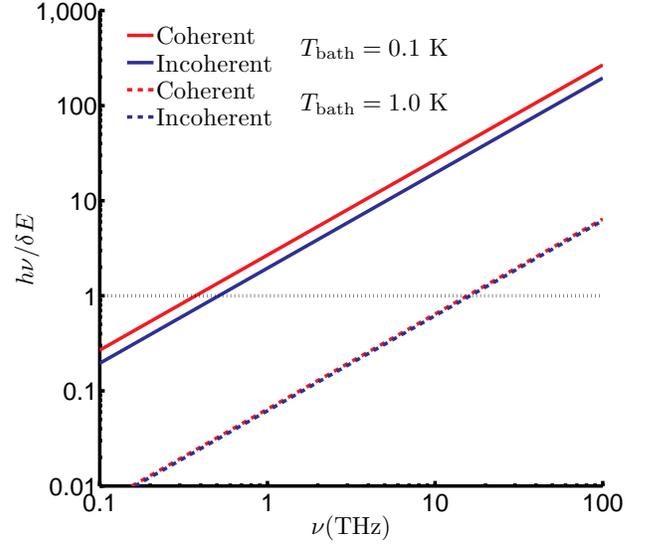}
\caption{(Color online) Detector resolving power for some operating temperatures. Note that the curves for $T_{\rm bath}=1.0$~K are nearly on top of each other.}\label{fig:resolution}
\end{figure}

The calorimetric mode is characterized by the resolving power
\[
 \frac{h\nu}{\delta E}=\frac{h\nu}{2\sqrt{2\,{\rm ln}\,2}{\rm NEP}_{\rm TFN}\sqrt{\tau}}.
\]
The estimate for the background noise, $\Delta E$, should be based on the idle state of the detector, i.e., on the state where $T_e=T_{\rm bath}$ and no signal is present. Here we rectify the erroneous statement made in Ref.~\onlinecite{KID}, where the noise is overestimated and the resolving power correspondingly underestimated.  We may use the approximative formulas given above for NEP$_{\rm TFN}$ and $\tau$ to obtain
\begin{equation}\label{eq:res}
\begin{split}
 \frac{h\nu}{\delta E}&=\frac{h\nu}{4\sqrt{2\,{\rm ln}\,2\,k_BT_{\rm bath}^2G_{\rm therm}}\sqrt\tau}\\
&=\frac{h\nu}{4\sqrt{2\,{\rm ln}\,2\,k_BT_{\rm bath}^2C_e}}\\
&=\frac{h\nu}{8\pi}\sqrt{\frac{3}{{\rm ln}\,2\,N_F\Omega k_B^3T_{\rm bath}^3}}.
\end{split}
\end{equation}
The scaling with $T_{\rm bath}^{-3/2}$ is approximatively valid for the temperature scales of our interest, where $C_e$ is nearly linear. Energy resolution of unity is obtained when
\begin{equation}\label{eq:nulimit}
 \nu=8\pi\sqrt{{\rm ln}\,2\,N_F\Omega k_B^3 T_{\rm bath}^3/3h^2}\approx 16\,{\rm THz}\times \tilde T_{\rm bath}^{3/2},
\end{equation}
and $\tilde T_{\rm bath}$ is the bath temperature in units of Kelvin. This means that at 100~mK, it should be possible to realize a single-photon detector for frequencies above 500~GHz. At 50~mK, this limit is 180~GHz. In contrast to the bolometer case, there is some freedom to alter the numerical prefactor in Eq.~\eqref{eq:nulimit} by choosing different material parameters from those given in the beginning of Sec.~\ref{sec:oper}. Since the proximity effect modifications in $\tau$ and NEP are in opposite directions, the resolving power is almost the same as that given in Eq.~\eqref{eq:res}, which neglects the proximity corrections. This is shown in Fig.~\ref{fig:resolution}.

\subsection{Optimization}\label{sec:opt}
The scheme described above is largely adjustable and such adjustment is required whenever the objective of the detector operation is changed, for example in the case of different power loads $P_T$ or operating temperatures $T_{\rm bath}$. While the exact optimal parameters depend largely on the purpose and the situation where the detector is used, we now provide some of the most interesting examples of the effects of altering the exact detector configuration and materials.

\subsubsection{Length}
The most notable material-independent characteristic of the detector is the absorber length $l$. This affects the Thouless energy $E_{\rm Th}\sim l^{-2}$ and, consequently, also the relation between $I_c$ and $T_e$, as shown in Fig.~\ref{fig:IcT} and the corresponding Eq.~\eqref{eq:Ic}. Since the detector is based on the strong observable response in $I_c$, which again is due to changes in $T_e$, altering $E_{\rm Th}$ markedly shifts the electron temperature regime where the detector operates. In Sec.~\ref{sec:bolo} it is shown that in bolometric operation it is either the intrinsic temperature fluctuations or the incident power itself that governs the noise level. Therefore it becomes crucial to be able to find the most appropriate operating temperature regime.

The operating regime is determined for two parameters $T_e$ and $T_{\rm bath}$ so that $T_{\rm min}\leq T_{\rm bath}<T_e\leq T_{\rm max}$ for the lower and the upper limits of the regime, $T_{\min}$ and $T_{\rm max}$.
In Sec.~\ref{sec:deta}, we determine the lower limit of operation to be $k_BT_{\rm min}=1.0E_{\rm Th}$ since below this temperature, $I_c$ saturates. The upper limit is set by our approximation of the minimum measurable critical current 3~nA. For our base value of absorber length, $l=1\,\mu$m, this corresponds to $T_{\rm max}=20E_{\rm Th}=400$~mK. The change in length has two effects. When $E_{\rm Th}$ becomes smaller, the curve in Fig.~\ref{fig:IcT} shifts left and the response diminishes, and vice versa. On one hand, long absorbers thus imply reduced supercurrent but on the other hand, this also means that $E_{\rm Th}$ is reduced and the detector can be operated at lower temperatures due to increased response in this regime. Conversely, if the absorber is short, the electron operating regime of the detector is shifted to higher temperatures and the critical current is increased for an easier readout.

When $l$ is \emph{increased} by a factor of 2, $k_BT_{\rm min}=E_{\rm Th}\rightarrow\frac 14 E_{\rm Th}\approx 5$~mK. Due to the decreasing supercurrent, we now have $T_{\rm max}\approx 55$~mK. These temperatures are already too small for present-day applicability. On the other hand, when $l$ is \emph{decreased} by a factor of 2, $k_BT_{\rm min}=E_{\rm Th}\rightarrow 4 E_{\rm Th}\approx 76$~mK and $T_{\rm max}\approx 2.4$~K. The drawback of this large temperature range is that we can expect the outdiffusion of energy to increase for the short absorber. This would change the fraction of the radiation power $r(\nu)$ which yields a thermal signal as discussed in Sec.~\ref{sec:rad}. We assess how changing of the absorber length affects $r(\nu)$ by noting that the only term in Eq.~\eqref{eq:processes} which scales with $l$ is the diffusion part proportional to $\tau_D^{-1}\sim l^2$. Changing $l$ therefore only changes the relative strength of diffusion compared to the irradiation and inelastic relaxation. The contribution to $r(\omega)$ due to diffusion comes from energies close to the gap $\Delta$, as seen in Fig.~\ref{fig:integrals}. However, in our parameter regime the effect of diffusion in Eq.~\eqref{eq:processes} is so weak compared to the inelastic energy re-distribution $I_{\rm e-e}$ and $I_{\rm e-ph}$ that a factor of 2 change in length of the absorber results only in a change of the order of 10-20 percent in $r$ at low frequencies $h\nu\lesssim 3\Delta$. At large frequencies, the energy redistribution through inelastic collisions is the dominant mechanism of power loss. We therefore conclude that the amount of radiation power that is coupled to the sensor is relatively independent of small changes in $l$ in our parameter regime.

\subsubsection{Graphene}
In Sec.~\ref{sec:bolo}, we have also determined that reducing the volume of the absorbing element, the intrinsic thermal noise can be decreased (see Eq.~\eqref{eq:NEP}). Consequently, graphene, which is composed of a single layer of graphite, can be seen as an attractive alternative for the absorber material when extremely weak radiation signals such as single photons are detected and $P_{\rm opt}$ is small. It has already been demonstrated that graphene, placed between superconductors, can support large supercurrents\cite{grapheneIS1, grapheneIS2} For graphene, the heat flow between the electrons and phonons does not follow Eq.~\eqref{eq:Q3D} anymore since this formula is valid only for 3D conductors. In Ref.~\onlinecite{grapheneP}, the electron-phonon cooling power in graphene has been determined at temperatures with a minimum of $T_e=20$~K. While this is already far above our operating regime, we may calculate $\dot Q_{\rm e-ph}$ for our metallic wires at such temperatures to get some idea where graphene stands when these two different realizations are compared.

According to Ref.~\onlinecite{grapheneIS2}, a graphene flake of width $w=9\,\mu$m and length $l=350$~nm with gate voltage such that carrier density becomes $n\sim 10^{12}\ldots 10^{13}$~1/cm$^2$ has prominent characters such as diffusion coefficient comparable to the values of our example above. More importantly, for such gating it has normal-state resistance $R_N\sim 50\,\Omega$ and could thus be used potentially as an absorber with good impedance matching. Since the length of the graphene absorber is reduced to one third of the value used for our metallic absorber, $E_{\rm Th}\sim l^{-2}$ is correspondingly increased by a factor of 9 and the operating regime in temperature shifts from 20~mK$\ldots$~400~mK to 200~mK$\ldots$~3.5~K. We use these specifications for the graphene detector. Then, from Ref.~\onlinecite{grapheneP}, the electron-phonon cooling power becomes 300~nW$\ldots3\,\mu$W ($n\sim 10^{12}\ldots 10^{13}$~1/cm$^2$). This should be compared with the value we get for the electron-phonon cooling power when we use Eq.~\eqref{eq:Q3D} for our metallic conductor at $T_e=20$~K and $T_{\rm bath}=0$. By extrapolating outside the operating regime of the metallic detector this way, $\dot Q_{\rm e-ph}\approx 1.6\,\mu$W. This is in the middle of the values we have for graphene, and we conclude that graphene can be used as a tunable detecting element with gate voltage determining the strength of relaxation. It is difficult to estimate what happens when the temperature is around $T_e=1$~K since direct extrapolation from the results of Ref.~\onlinecite{grapheneP} would require knowing the exact temperature dependence of $\dot Q_{\rm e-ph}$ for graphene. However, we expect that by adjusting the carrier density in graphene by the gate voltage, the thermal conductance of the electron-phonon link can be made small in comparison to the case of the metallic wire, thus producing favourable NEP figures.

\section{Conclusion}
In this paper, we present in detail the operating concept of the proximity Josephson sensor introduced in Ref.~\onlinecite{KID} and study the physical processes involved in the detector operation. First, we devise a model to explain how the incident radiation couples to the sensor, how the radiation energy is distributed inside the absorber after the initial excitation and how the energy finally escapes the system. We adopt the hot-electron model where the absorption of radiation results in increase of the electron temperature. Then we show how the superconducting proximity effect in the absorbing element of the detector alters the heat capacity and the heat conductance of the detector and how this affects the detector performance in terms of the nonequilibrium noise. We observe that in the operating regime of the detector, where temperature is larger than the Thouless energy, corrections to the expected magnitude of noise due to the proximity effect are small. Nevertheless, they reduce the intrinsic thermal noise of the detector, which sets the limit for detector performance.

Since the detector is largely adjustable depending on the operating temperature and power load, we refrain from giving any definite limits of performance. However, we show the physical limit of minimum noise for a given set of material parameters in terms of the base NEP in Fig.~\ref{fig:NEP}. This base NEP sets the ultimate limit for single-photon detection in the calorimetric mode and in the bolometric mode it sets the limit for small input powers. For large input powers, this NEP is to be compared with the NEP arising from the heating, given in Eq.~\eqref{eq:powerNEP}. Notably, the latter is only weakly dependent of the volume or the strength of electron-phonon coupling in the absorber. We observe that at $T\sim 100$~mK, base NEP of the order of $10^{-20}$~W/$\sqrt{\rm Hz}$ can be expected. Under power loads of several picowatts, corresponding to a temperature raise to a few Kelvin, NEP becomes $\sim 10^{-17}$~W/$\sqrt{\rm Hz}$. For power loads this large it is desirable to fabricate devices with larger volumes, leading to smaller temperature changes and, consequently, lower NEP. It follows that when designing a detector for a particular purpose using the PJS concept, we need to balance between the intrinsic (base) noise of the absorber at $T_e=T_{\rm bath}$ and the noise that results from heating the absorber above $T_{\rm bath}$.

The dynamic range of the detector can be of the order of $10^6$ but if the detector is operated at a large range of input powers, the nonlinearity in response needs to be taken into account. On the other  hand, the numbers for detector resolving power suggest that our scheme could also be used effectively in detection of extremely weak radiation signals. This includes single-photon detection for frequencies above 180~GHz and even detailed radiation spectroscopy at the terahertz regime. Optimizing the detector for a specific task requires choosing appropriate materials, and for this we have studied the behavior of graphene as the absorber. Notably, both the graphene and metallic wire concepts share the advantage that, by controlling the length-to-width ratio of the absorber, a good impedance matching to the antenna circuit is possible. However, we encounter additional mechanisms of power loss in optical coupling of the incident power to the absorber, namely the outdiffusion of quasiparticles above the superconducting energy gap and nonthermal redistribution of radiation-excited quasiparticles. In a typical range of materials parameters, the latter is so dominant that the length of the absorber may be varied without experiencing a drastic increase in outdiffusion allowing the detector to be optimized for varying operating requirements. This is especially important if the detector is to be operated at temperatures over 1~Kelvin. In that case, it needs to be investigated whether sub-micron or graphene absorbers may be used.

\begin{acknowledgments}
The authors thank Francesco Giazotto, Pertti Hakonen, Panu Helist\"o, Leonid Kuzmin, Arttu Luukanen, and Andrei Zaikin for useful discussions. JV is supported by the Finnish Foundation for Technology Promotion whereas TTH acknowledges the support from the Academy of Finland and MAL from the Finnish Academy of Science and Letters.
\end{acknowledgments}

\end{document}